\documentclass[12pt]{article}
\usepackage{graphics}
\usepackage{epsfig}
\begin{document}
\title{{Nuclear parton distribution functions and  energy loss effect
 in the Drell-Yan reaction off nuclei  }
%\thanks {Supported by  Natural
%Science Foundation of China(90103020,10075057,10175074) , CAS
%Knowledge Innovation Project(KJCX2-SW-N02),Major State Basic
%Research Development Program (G20000774),  Natural Science
%Foundation of Hebei Province(103143)}
}
\author{ C.-G. Duan $^{1,2,5}$\thanks{E-mail:duancg$@$mail.hebtu.edu.cn}
L.-H. Song $^3$  \\
S.-H. Wang  $^{4}$ G.-L. Li $^{2,5}$
 \\
{\small 1.Department of Physics, Hebei Normal University, Shijiazhuang,050016,P.R.China}\\
{\small 2.Institute of High Energy Physics,CAS,
          Beijing,100039,P.R.China}\\
{\small 3.College of Information, Hebei Polytechnic University,Tangshan,063009,P.R.China}\\
{\small 4.Electrical Engineering Department, Shijiazhuang Railway
          Institute,Shijiazhuang,050043,P.R.China}\\
{\small
5.CCAST(WorldLaboratory).P.O.Box8730,Beijing,100080,China}}

\date{}
\maketitle
\baselineskip 9mm
\vskip 0.5cm
\begin{center}
\begin{minipage}{134mm}
\begin{center}
Abstract
\end{center}
The energy loss effect in  nuclear matter is another nuclear
effect apart from the nuclear effects on the parton distribution
as in deep inelastic scattering process. The quark energy loss can
be measured best by the nuclear dependence of the high energy
nuclear Drell-Yan process. By means of two typical kinds of quark
energy loss parametrization and the different sets of nuclear
parton distribution functions, we present a analysis of the E866
experiments on the nuclear dependence of Drell-Yan lepton pair
production resulting from the bombardment of Be, Fe and W targets
by 800GeV protons at Fermilab. It is found that the quark energy
loss in cold nuclei is strongly dependent on the used nuclear
parton distribution functions. The further prospects of using
relatively low energy proton incident on  nuclear targets  are
presented  by combining the quark energy loss rate determined from
a fit to the E866 nuclear-dependent ratios versus $x_1$, with the
nuclear parton distribution functions given from  lA deep
inelastic scattering (DIS) data. The experimental study of the
relatively low energy nuclear Drell-Yan process can give valuable
insight in  the enengy loss of fast quark propagating a cold
nuclei and help to pin down nuclear parton distributions
functions.  


\end{minipage}
\end{center}

\vskip 0.5cm

{\bf I Introduction }

\hspace{0.5cm}In proton-proton collisions, we learn about the
interactions between the quarks and gluons that make up the
colliding nucleon$^{[1]}$. Parton distribution functions in
nucleon have been obtained by the relative high-energy reaction
data $^{[2]}$. These analysis help us calculate precise cross
sections for finding new physics phenomena. In nucleus-nucleus
collisions, we may find a signal for the existence of the
deconfining phase of QCD, the quark-gluon plasma $^{[3]}$. In
proton-nucleus collisions, we hope to gain information about the
modification of the parton distribution functions in nucleon when
it is immersed in the nuclei, and to learn the space-time
development of the strong interaction during its early stages.
Understanding the initial stages of ultrarelativistic heavy ion
collisions is of utmost importance in order to understand the
outcome of the high energy heavy ion experiments, such as the BNL
relativistic heavy ion collider(RHIC) and CERN large hadron
collider (LHC). Understanding the modifications of the parton
distribution functions  and the parton energy loss in nuclei
should be the first important step towards pinning down the
initial conditions of a heavy-ion collision and understanding of
$J/\psi$ production which is required if it is to be used as a
signal for the quark-gluon plasma in relativistic heavy ion
collisions.

\hspace{0.5cm}The production of lepton pairs in proton-nucleus
collisions, the Drell-Yan process$^{[4]}$, is   one of most
powerful tools to probe the structure of  nuclei, and the
propagating of partons through cold nuclei. Its parton model
interpretation is straightforward
--- the process is induced by the annihilation of a
quark-antiquark pair into a virtual photon which subsequently
decays into a lepton pair. The Drell-Yan process in proton-nucleus
collisions therefore is closely related to the  quark distribution
functions in nuclei. Unlike DIS, it is directly sensitive to
antiquark contributions in target parton distributions. When DIS
on nuclei occurs at $x<0.08$, where x is the parton momentum
fraction, the cross section per nucleon decreases with increasing
nucleon number A due to shadowing$^{[5]}$. Shadowing should also
occur in Drell-Yan dimuon production at small $x_2$, the momentum
fraction of the target parton, and theoretical calculations
indicate that shadowing in the DIS and Drell-Yan reactions has a
common origin $^{[6]}$.

\hspace{0.5cm} In high energy inelastic hadron-nucleus scattering,
the projectile rarely retains a major fraction of its momentum
after traversing the nucleus. Rather, its momentum is shared by
several produced particles, which form a hadron jet in the forward
direction. The classical description of this phenomena is that the
projectile suffers multiple collisions and repeated energy loss in
the nuclear matter. In other words, each quark or gluon in the
projectile can loss a finite fraction of its energy in the nuclear
target due to QCD bremsstrahlung$^{[7]}$. The Drell-Yan
reaction$^{[4]}$ on nuclear targets provides, in particular, the
possibility of probing the propagation of quark through nuclear
matter, with the produced lepton pair carrying away the desired
information on the projectile quark after it has travelled in the
nucleus. Only initial-state interactions are important in
Drell-Yan process since the dimuon in the final state does not
interact strongly with the partons in the nuclei. This makes
Drell-Yan scattering an ideal tool to study energy loss.
Therefore, shadowing and initial state partonic energy loss are
processes that occur in  the proton-induced Drell-Yan reaction on
nuclei.

\hspace{0.5cm}In order to describe the modification of the initial
state parton distributions in nucleus, a  variety of approach to
this question exist in the literature$^{[8]}$. Recently, there are
two groups doing global analysis of nuclear parton distribution
functions. Eskola, Kolhinen, Ruuskanen and Salgado(EKRS) produces
EKS98 package of nuclear parton distributions$^{[9]}$. Hirai,
Kumano, Miyama and Nagai derived several sets  of nuclear parton
distribution functions from extensive experimental
data$^{[10,11]}$. In 1999, Eskola, Kolhinen, Ruuskanen and
Salgado(EKRS)$^{[9]}$ suggested a set of nuclear parton
distributions, which are studied within a framework of the DGLAP
evolution. The measurements of $F^A_2/F^D_2$ in deep inelastic
$lA$ collisions, and Drell-Yan dilepton cross sections measured in
$pA$ collisions were used as constraints. The kinematical ranges
are $10^{-6}\leq x\leq 1$  and $2.25GeV^2\leq
Q^2\leq10^{4}GeV^{2}$  for nuclei from deuteron to heavy ones.
With the nuclear parton distributions, the calculated results
agreed very well with the relative EMC and Fermilab E772
experimental data$^{[12]}$. In 2001, Hirai,Kumano and
Miyama(HKM01)$^{[10]}$ proposed two types  of nuclear parton
distributions which were obtained by quadratic and cubic type
analysis, and determined by a $\chi^2 $ global analysis of
existing experimental data on nuclear structure functions without
including the proton-nucleus Drell-Yan process. The kinematical
ranges covered $10^{-9}\leq x\leq 1$  and
 $1GeV^2\leq Q^2\leq10^{5}GeV^{2}$  for nuclei from deuteron to heavy
ones. As a result, they obtained reasonable fit to the measured
experimental data of $F_2$. In 2004, Hirai,Kumano and
Nagai(HKN04)$^{[11]}$ re-analyze experimental data of nuclear
structure function ratios $F^{A}_{2}/F^{A'}_{2}$ and Drell-Yan
cross section ratios for obtaining their another parton
distribution functions in nuclei. In HKN04, Drell-Yan data
$^{[12,13]}$ are included for determining the sea quark
modification in the range $ 0.02<x_2<0.2$. In addition, HERMES
data$^{[14]}$ are used. In this work, we will use these
parameterizations and investigate the nuclear dependence of the
Drell-Yan process.

\hspace{0.5cm}Fermilab Experiment866(E866) $^{[13]}$ performed the
precise measurement of the ratios of the Drell-Yan cross section
per nucleon for an 800GeV proton beam incident on Be, Fe and W
target at larger values of $x_1$, the momentum fraction of the
beam parton, larger values of $x_F$($\approx x_1-x_2$), and
smaller values of $x_2$ than reached by the previous experiment,
Fermilab E772$^{[12]}$. The extended kinematic coverage of E866
significantly increases its sensitivity to energy loss and
shadowing. This is the first experiment on the energy loss of
quark passing through a cold nucleus.

\hspace{0.5cm}For many years it has been suggested that fast quark
energy loss might give rise to a nuclear dependence$^{[15,16,17]}$
of the cross section of Drell-Yan. After the E866 experimental
data was reported,several groups have given their theoretical
analysis of the data$^{[18,19,20]}$. In previous report$^{[20]}$,
by means of EKRS and HKM01  nuclear parton distribution functions,
we investigated the Drell-Yan production cross section ratios from
E866 data  in the framework of Glauber model. We found that the
theoretical results with energy loss are in good agreement with
the Fermilab E866 experiment by means of HKM01 nuclear parton
distributions. However, the calculated results without energy loss
can give good fits by using  EKRS nuclear parton distribution
functions. In this report, the nuclear dependence of the pA
Drell-Yan production cross sections are studied  by combining two
typical kinds of  quark energy loss parametrization  with the EKRS
, HKM01 and HKN04  nuclear parton distribution. Using the values
of quark energy loss from a fit to E866 experimental data, the
prospects are given for the lower energy proton beams  off
deuteron and tungsten. Comparing with future experiments can give
valuable insight in the enengy loss of fast quark propagating a
cold nuclei and help to pin down nuclear parton distributions
 functions.

{\bf II  Nuclear Drell-Yan reaction }

\hspace{0.5cm}In the Drell-Yan process$^{[4]}$, the leading-order
contribution is quark-antiquark annihilation into a lepton pair.
The annihilation cross section can be obtained from the
$e^{+}e^{-}\rightarrow\mu^{+}\mu^{-}$ cross section by including
the color factor $\frac{1}{3}$ with the charge $e^{2}_{f}$ for the
quark of flavor $f$.
\begin{equation}
   \frac{d\hat{\sigma}}{dM}=\frac{8\pi\alpha^2}{9M}e^2_f\delta(\hat{s}-M^2),
\end{equation}
where $\sqrt{\hat{s}}=(x_1x_2s)^{1/2}$, is the center of mass
system (c.m.system) energy of $q\bar{q}$ collision,
$x_1$(resp.$x_2$)is the momentum fraction carried by the
projectile (resp.target) parton, $\sqrt{s}$ is the center of mass
energy of the hadronic collision, and $M$ is the invariant mass of
the produced dimuon. The hadronic Drell-Yan differential cross
section is then obtained from the convolution of the above
partonic cross section with the quark distributions in the beam
and in the target :
\begin{equation}
 \frac{d^2\sigma}{dx_1dM}=K\frac{8\pi\alpha^2}{9M}\frac{1}{x_1s}
 \sum_{f}e^2_f[q^p_f(x_1)\bar{q}^A_f(x_2)
 +\bar{q}^p_f(x_1)q^A_f(x_2)],
\end{equation}
where $ K$ is the high-order QCD correction, $\alpha$ is the
fine-structure constant, the sum is carried out over the light
flavor $f=u,d,s$, and $q^{p(A)}_{f}(x)$ and ${\bar
q}^{p(A)}_{f}(x)$ are the quark and anti-quark distributions in
the proton (nucleon in the nucleus A). In order to obtain the
$x_1$ dependence of Drell-Yan production, we shall  deal in the
following with the single differential cross section,
\begin{equation}
 \frac{d\sigma}{dx_1}=K\frac{8\pi\alpha^2}{9x_1s}
 \sum_{f}e^2_f\int\frac{dM}{M}[q^p_f(x_1)\bar{q}^A_f(x_2)
 +\bar{q}^p_f(x_1)q^A_f(x_2)],
\end{equation}
where the integration over the dimuon mass is performed in the
range given from E866 experiment.

\hspace{0.5cm}Now let us take into account of the energy loss of
the fast quarks moving through the cold nuclei.In this work, we
will introduce two typical kinds of quark energy loss expressions.
One is given by Brodsky and Hoyer$^{[7]}$ from uncertainty
principle, $\Delta x_1 \propto A^{1/3}$, which can be rewritten as
\begin{equation}
\Delta x_1= {\alpha}\frac{<L>_A}{E_p},
\end{equation}
where $\alpha$ indicate the  incident quark  an energy loss per
unit length in nuclear matter, $<L>_A$ is the average path length
of the incident quark in the nucleus A, $E_p$ is the energy of the
incident proton. The average path length is employed using the
conventional value, $<L>_A=3/4(1.2A^{1/3)}$fm $^{[22]}$. In
addition to the linear quark energy loss rate, another is deduced
by Baier et.al.$^{[21]}$as$\Delta x_1 \propto A^{2/3}$, which can
be rewritten as
\begin{equation}
\Delta x_1= {\beta}\frac{<L>^2_A}{E_p}.
\end{equation}
Obviously, the partonic energy loss is quadratic with the path
length.

\hspace{0.5cm}After considering the  quark energy loss in nuclei,
the incident quark momentum fraction can be shifted from
$x'_1=x_1+\Delta x_1$ to $x_1$ at the point of fusion. Combining
the shadowing with initial state energy loss, the production cross
section in pA Drell-Yan process can be written as
\begin{equation}
 \frac{d\sigma}{dx_1}=K\frac{8\pi\alpha^2}{9x_1s}
 \sum_{f}e^2_f\int\frac{dM}{M}[q^p_f(x'_1)\bar{q}^A_f(x_2)
 +\bar{q}^p_f(x'_1)q^A_f(x_2)].
\end{equation}

{\bf III  Constraint on quark energy loss from E866 }

\hspace{0.5cm}In order to pin down quark energy loss by comparing
with the experimental data from E866 collaboration$^{[13]}$, we
introduce the nuclear Drell-Yan ratios as:
\begin{equation}
R_{A_{1}/A_{2}}(x_{1})=\frac{d\sigma^{p-A_{1}}}{dx_ {1}}/{\frac
{d\sigma^{p-A_{2}}}{dx_{1}}}.
\end{equation}
The integral range on M is determined according to the E866
experimental kinematic region. In our theoretical analysis,
$\chi^2$ is calculated with the Drell-Yan differential cross
section rations $R_{A_1/A_2}$ as
\begin{equation}
\chi^2=\sum\limits_{j}\frac{(R^{data}_{A_1/A_2,j}-R^{theo}_{A_1/A_2,j})^2}
{(R^{err}_{A_1/A_2,j})^2},
\end{equation}
where the experimental error is given by systematic errors as
$R^{err}_{A_1/A_2,j} $, and $ R^{data}_{A_1/A_2,j}$(
$R^{theo}_{A_1/A_2,j}$) indicates the experimental data
(theoretical values ) for the ratio $R_{A_{1}/A_{2}}$.

\hspace{0.5cm} Taking advantage of the EKRS$^{[9]}$ nuclear parton
distribution functions with Eq.(4), the obtained $\chi^2$ value is
$\chi^2=51.4$ for the 56 total data points when
$\alpha=0.0$(without energy loss effects). The $\chi^2$ per
degrees of freedom is given by $\chi^2/d.o.f.=0.918$. It is
apparent that theoretical results without energy loss effects
agree very well with the E866 experimental data. We consider also
combining HKM01 cubic type of nuclear parton distribution$^{[10]}$
with the linear quark energy loss parameterizations,i.e.Eq.(4).
With $\alpha=0.0$(without energy loss effects), the obtained
$\chi^2$ per degrees of freedom is $\chi^2/d.o.f.=2.526$. With
$\alpha=1.99$(with energy loss effects), the obtained  $\chi^2$
per degrees of freedom are $\chi^2/d.o.f.=1.008$.  The results
given by HKM01 quadratic type are nearly the same as these above.
As an example, the calculated results with energy loss expression
 are shown in Fig.1 and Fig.2. which is the Drell-Yan cross
section ratios for Fe  to Be and W to Be as functions of $x_1$ for
various interval of $M$, respectively. The solid curves are the
ratios with only the nuclear effect on the parton distribution as
in DIS scattering process, and the dotted curves correspond to an
energy loss effect  with nuclear effect on structure function.
From comparison with the experimental data, it is found that our
theoretical results with energy loss effect are in good agreement
with the Fermilab E866. If employing the HKN04 nuclear parton
distribution function $^{[11]}$, With $\alpha=0.0$(without energy
loss effects), the obtained $\chi^2$ per degrees of freedom is
$\chi^2/d.o.f.=2.526$. With $\alpha=1.92$(with energy loss
effects), the obtained  $\chi^2$ per degrees of freedom are
$\chi^2/d.o.f.=1.045$. It is obvious that the results with HKM01
are most near to those with HKN04. We notice that HKM01 don't use
the nuclear Drell-Yan data, and HKN04 include  E772 and E866
Drell-Yan experimental data.
\tabcolsep0.5cm
\begin{table}
\caption{The results in detail from a fit to E866 with HKM01 and
HKN04}
\begin{center}
\begin{tabular}{|c|c|c|}
\hline
         & $\alpha=1.99$( HKM01)    & $\alpha=1.92$( HKN04)\\
\hline

     $\chi^2/d.o.f.$(Fe/Be)&  0.873 & 0.898\\
\hline
      $\chi^2/d.o.f.$(W/Be)  & 1.143 & 1.193\\
\hline

\end{tabular}
\end{center}
\end{table}
We give the results from a fit to E866 $R_{W/Be}$ and $R_{Fe/Be}$
in Table 1. It can be seen from the Table that they are similar by
means of HKM01 and HKN04. Although HKN04 include the E772 and E866
Drell-Yan cross section ratios versus $x_2$, HKN04 don't give a
good fit to W/Be Drell-Yan ratios at small $x_2$$^{[11]}$, which
may be the reason for two similar results.

\hspace{0.5cm}In Fig.3, we show the nuclear modifications of sea
quark distributions in EKRS(dotted line), HKM01(dashed line) and
HKN04(solid line) at $Q^2=5.0GeV^2$ for Be/D(up), Fe/D(middle) and
W/D(down). It is found that the trend is the same for EKRS, HKM01
and HKN04 in the region $x_2<0.12$. The differences occur in the
region $x_2>0.12$ among EKRS, HKM01 and HKN04. For E866 Drell-Yan
measurement, the kinematic ranges cover $0.01<x_2<0.12$ and
$0.21<x_1<0.95$ with dimuon mass in the range $4.0<M<8.4GeV$.
Therefore, the results from HKM01($\alpha=1.99$) are similar to
those of HKN04($\alpha=1.92$). The sea quark modifications in EKRS
is the lowest one, so that we obtain the theoretical results
without quark energy loss in good agreement with the E866
experimental data. It is noticeable that HKN04 employ E772 and
E866 Drell-Yan data, and EKRS include E772 experimental data.

{\bf IV  Prospects and summary }

\hspace{0.5cm}It is demonstrated that the effects of quark energy
loss are largest at lower incident proton energies at larger
$x_1$$^{[22]}$. In the future, the Fermilab Main Injector(FMI,
120GeV proton beam)$^{[23]}$ and the  Japan Proton Accelerator
Research Complex (J-PARC, 50GeV proton beam)$^{[24]}$, where
shadowing effect disappears and energy loss effect of fast quarks
could provide the dominant nuclear dependence, will be operated.
The precise measurements of the nuclear dependent of Drell-Yan
production can shed light on the quark energy loss. The HKM01
cubic type of nuclear parton distribution functions are employed
in the following discussion. Figure 4 shows how quark energy loss
would affect the $(p+W)/(p+D)$ per nucleon Drell-Yan cross
sections at 50GeV and 120GeV proton beam. The kinematic ranges
cover $M>4.2GeV$ in order to avoid contamination from charmonium
decays. In this calculation, the energy loss per unit length
$\alpha=1.99GeV/fm$ from a good fit to E866 with HKM01 nuclear
parton distribution functions.

\hspace{0.5cm} In addition, nuclear dependent Drell-Yan data can
also further determine whether this energy loss is linear or
quadratic with the path length. The $(p+W)/(p+D)$ per nucleon
Drell-Yan cross section ratios are given in Fig.5 where the solid
and dotted lines correspond to a quadratic energy loss of
$\beta=0.29GeV/fm^2$ and to a linear energy loss of
$\alpha=1.99GeV/fm$ from a fit to E866 at 120GeV and 50GeV proton
beams,respectively. As seen in Fig.5, we can easily distinguish
between $L$ and $L^2$ dependence of energy loss.

\hspace{0.5cm}Although there are currently abundant data on
electron and moun deep inelastic scattering off nuclei, it is
difficult to determine nuclear valence quark distributions in the
small $x$ region and the nuclear anti-quark distributions in the
$x>0.2$ region. Nuclear valence quark distributions in medium- and
large-x region can be relatively well determined. It is well
considered that the precise nuclear parton distributions must be
known in order to calculate cross sections of high energy nuclear
reactions accurately and find a signature of quark-gluon plasma in
high energy heavy-ion reactions. We suggest using precise neutrino
scattering experimental data, which can provide a good  method for
measuring the $F_2(x,Q^2)$ and $xF_3(x,Q^2)$ structure functions.
Using the average of $xF_3^{\nu A}(x,Q^2)$ and
$xF_3^{\bar{\nu}A}(x,Q^2)$, the nuclear valence quark distribution
functions can be well be clarified$^{[25,26]}$. The nuclear
antiquark distribution can fixed by means of  $F_2(x,Q^2)$ and
Drell-Yan experimental data. From our results,  the energy-loss
effects are large in the large-$x_1$ region, especially in
low-energy experiments (Fig.4). However, they are not large
effects at moderate  $x_1$, as shown in  Figs.1 and 2, in the
Fermilab experiments. In order to determine the nuclear antiquark
distribution in the region, $x>0.2$, we need another Drell-Yan
experiment at lower incident proton energies. We suggest that,
considering the existence of quark energy loss, the energy-loss
effects should be taken into account for the extraction of precise
nuclear parton distribution functions from the Drell-Yan
experimental data.

\hspace{0.5cm}In summary, we have made a leading-order analysis of
E866 data in nuclei  by taking into account of the energy loss
effect of fast quarks. Our theoretical results with quark energy
loss are in good agreement with the Fermilab E866 experiment by
means of the HKM01 and HKN04 parametrizations of nuclear parton
distributions, which is the same as that in our previous
work$^{[20]}$. We find that  the quark energy loss is close to
nuclear parton distribution functions. We desire to operate
precise measurements of the experimental study of the relatively
low energy nuclear Drell-Yan process. These new experimental data
can shed light on the enengy loss of fast quark propagating in a
cold nuclei and help to pin down nuclear parton distributions
functions which have a direct impact on the interpretation of many
hard scattering processes in nuclei.

{\bf Acknowledgement:}The authors thank J.C.Peng for useful
discussions by e-mail.This work is partially supported by Natural
Science Foundation of China(10175074) , CAS Knowledge Innovation
Project (KJCX2-SW-N02),Major State Basic Research Development
Program (G20000774),  Natural Science Foundation of Hebei
Province(103143)

%\vskip 2cm
%\begin{center} Figure caption
%\end{center}
%Fig.1 Fig.2 Fig.3 Fig.4 Fig.5

\newpage
\begin{figure}
\centering
\includegraphics[width=1.1\textwidth]{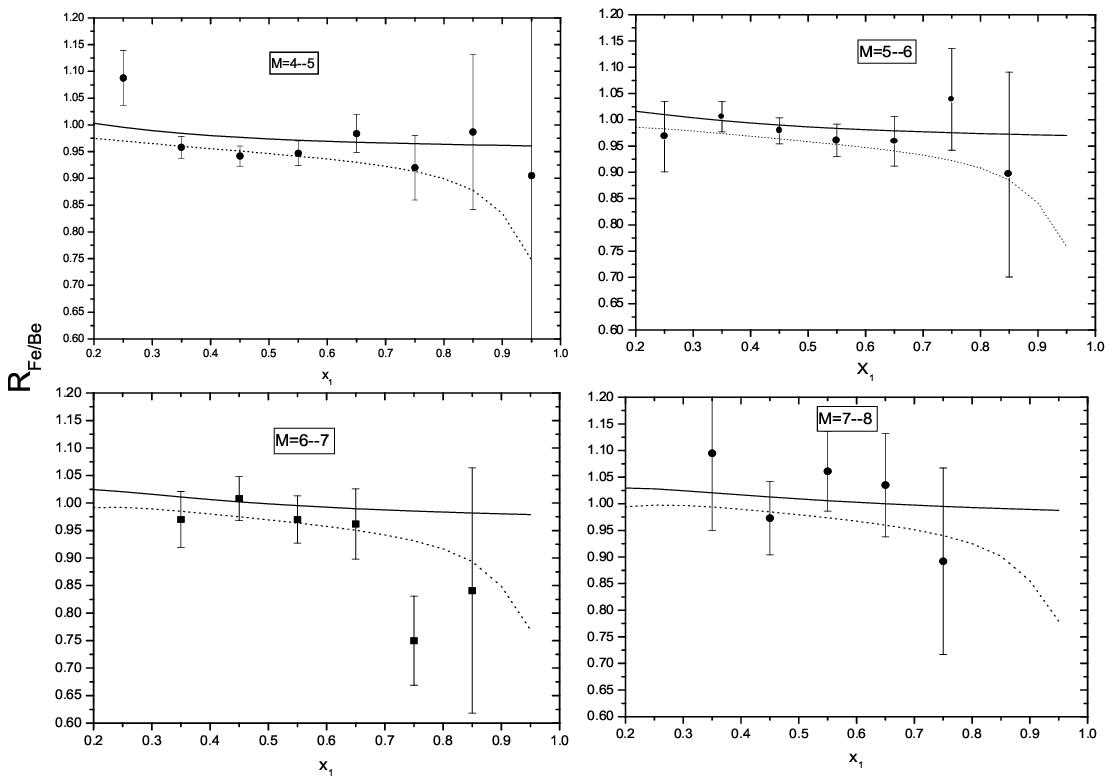}
\caption{The nuclear Drell-Yan cross section ratios
$R_{{A_1}/{A_2}}(x_1)$ on Fe to Be for various intervals M. Solid
curves correspond to nuclear effect on structure function. Dotted
curves  show the combination of shadowing and energy loss effect
with HKM01 cubic type of nuclear parton distributions. The
experimental data are taken from the E866[13].}
\end{figure}

\newpage
\begin{figure}
\centering
\includegraphics[width=1.1\textwidth]{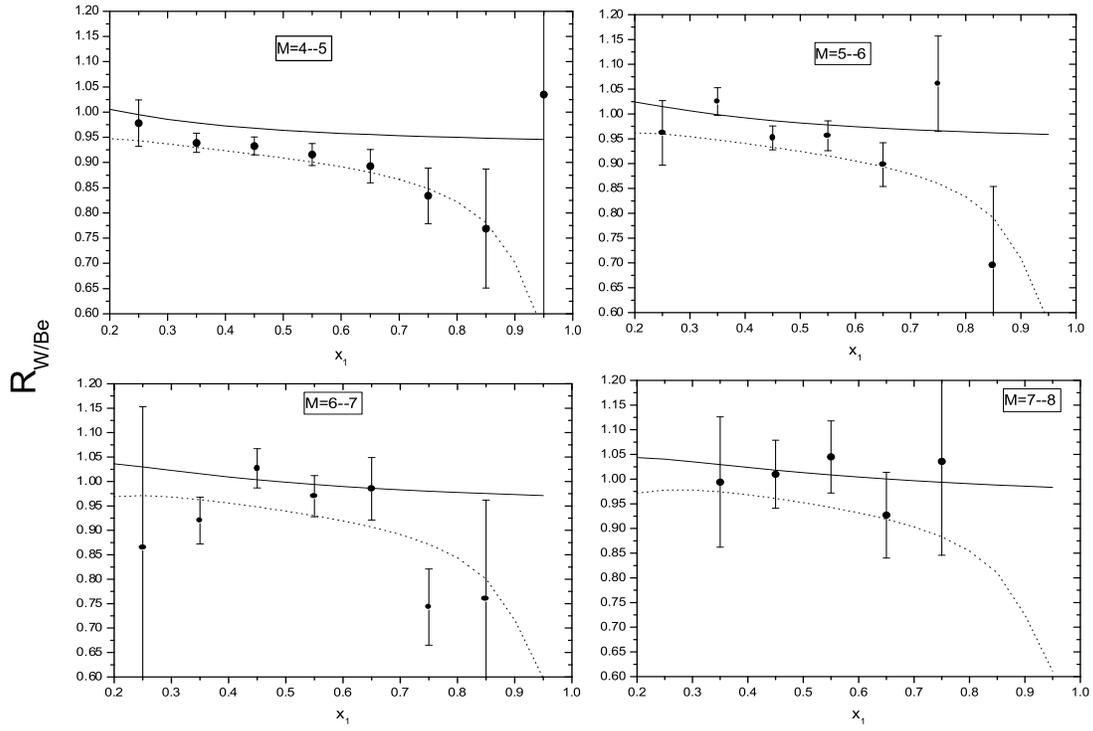}
\caption{The nuclear Drell-Yan cross section ratios
$R_{{A_1}/{A_2}}(x_1)$ on W to Be for various intervals M. The
comments are the same as Fig.1}
\end{figure}

\newpage
\begin{figure}
\centering
\includegraphics[width=0.9\textwidth]{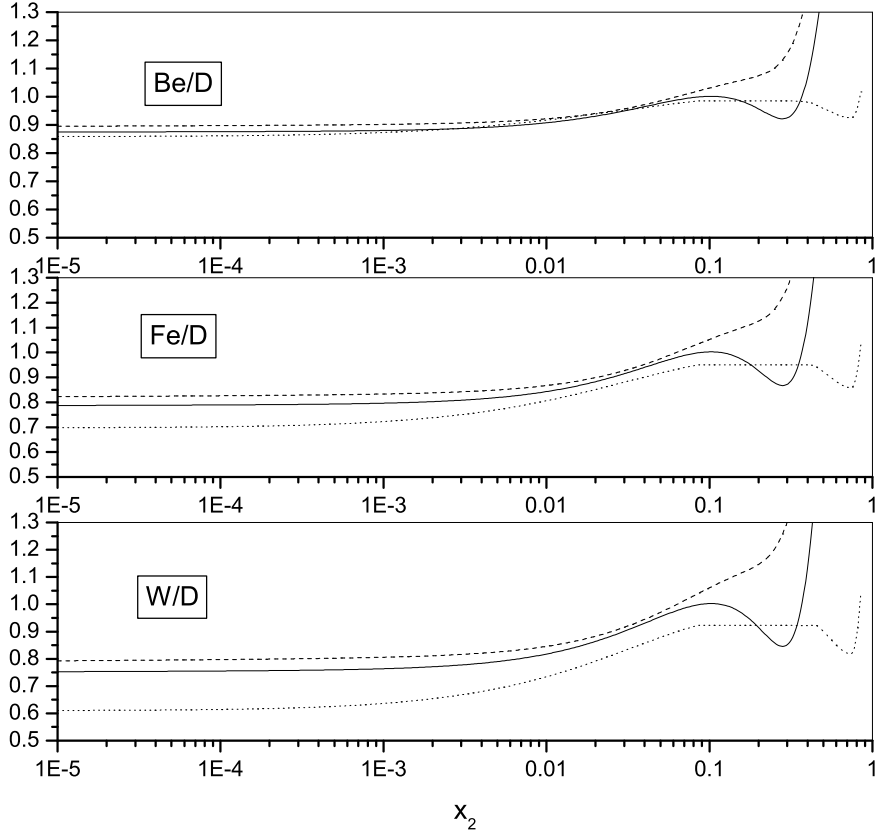}
\caption{The nuclear modifications of sea quark distributions in
EKRS(dotted line), HKM01(dashed line) and HKN04(solid line) at
$Q^2=5.0GeV^2$ for Be/D, Fe/D and W/D.}
\end{figure}

\newpage
\begin{figure}
\centering
\includegraphics[width=1.1\textwidth]{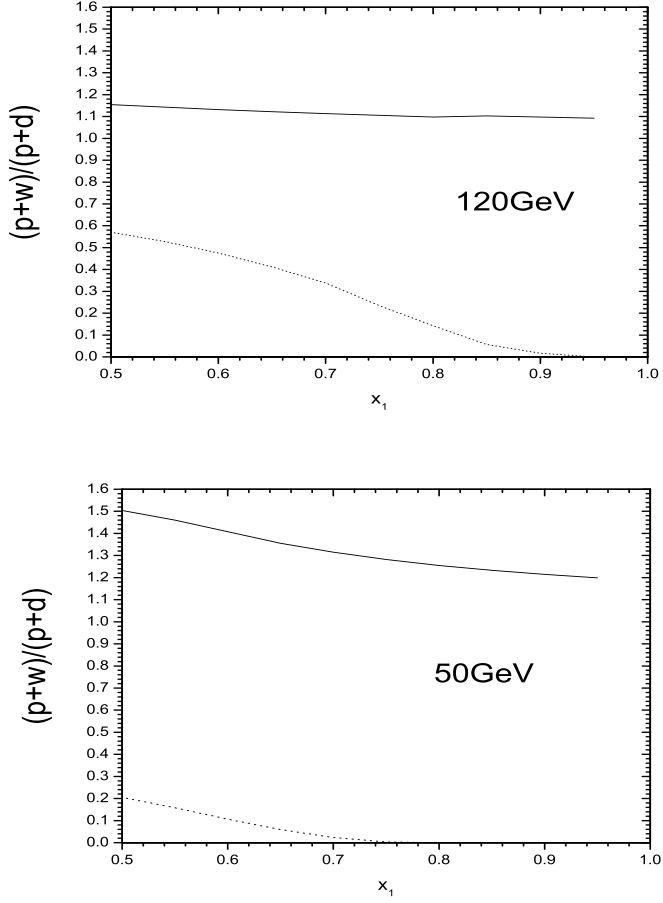}
\caption{The nuclear Drell-Yan cross section ratios
$R_{{A_1}/{A_2}}(x_1)$ on W to D at 120GeV and 50GeV incident
proton beams with a linear energy loss $\alpha=1.99GeV/fm$. Solid
curves correspond to nuclear effect on structure function. Dotted
curves  show the combination of shadowing and energy loss effect
with HKM01 cubic type of nuclear parton distributions.}
\end{figure}

\newpage
\begin{figure}
\centering
\includegraphics[width=1.1\textwidth]{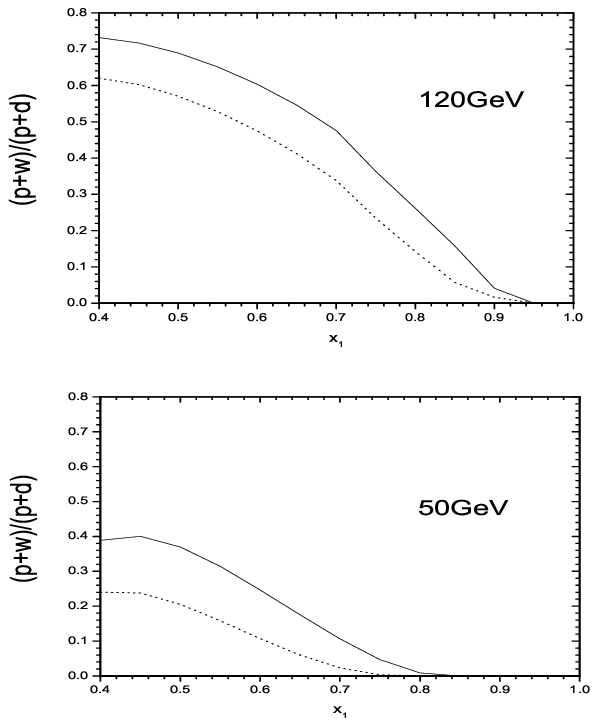}
\caption{The nuclear Drell-Yan cross section ratios
$R_{{A_1}/{A_2}}(x_1)$ on W to D at 120GeV and 50GeV incident
proton beams with a linear energy loss of $\alpha=1.99GeV/fm$ and
a quadratic energy loss of $\beta=0.29GeV/fm^2$. Solid curves
correspond to a quadratic energy loss. Dotted curves show the
linear energy  loss effect with HKM01 cubic type of nuclear parton
distributions.}
\end{figure}

%\newpage
%\begin{figure}[htb]
%\begin{center}

%\epsfig{file=Fig.eps,width=23cm}
%\caption{%%%%}
%\end{center}
%\end{figure}


\vskip 1cm
\begin{thebibliography}{s2}
\bibitem{s1}  H.Fritzsh,Phys.Lett.B67(1977)217;\\
              M.Gluck,J.F.Owens.and E.Reya,Phys.Rev.D17(1976)2324.
\bibitem{s2}  http://durpdg.drr.ac/hepdata/pdf.html
\bibitem{s3}  T.Matsui and H.Satz,Phys.lett.B178(1986)1986.
\bibitem{s4}  S.Drell and T.M.Yan, Phys.Rev,Lett.,25(1970)316.
\bibitem{s5}  M.Arneodo.et.al.(EMC), Nucl.Phys,B441(1995)3.
\bibitem{s6}   S.J.Brodsky,A.Hebecker and E.Quark,Phys.Rev.,D55(1997)2584.
\bibitem{s7}   S.J.Brodsky and P.Hoyer,Phys.lett.,B298(1993)165.
\bibitem{s8}   D.F.Geesaman,K.Saito and A.W.Thomas,Ann.Rev.Nucl.Part.Sci.45(1995)337
\bibitem{s9}   K.J.Eskola,V.J.Kolhinen and C.A.Salgado,Eur.Phys.J.C9(1999)61.\\
              K.J.Eskola,V.J.Kolhinen and P.V.Ruuskanen,Nucl.Phys.B535(1998)351.
\bibitem{s10} M.Hirai,S.Kumano,M.Miyama,Phys.Rev.D64(2001)034003.
\bibitem{s11} M.Hirai,S.Kumano,T.-H.Nagai,hep-ph/0404093.
\bibitem{s12} D.M.Adle et al.(E772),Phys.Rev.Lett.,64(1990)2479
\bibitem{s13} M.A.Vasiliev,et.al.(E866),Phys.Rev.Lett.83(1999)2304.
\bibitem{s14} A.Airapetian et.al.(HERMES),Phys.Lett.,B567(2003)339.
\bibitem{s15} G.T.Bodwin,S.J.Brodsky and G.P.Lepage,Phys.Rev.,D39(1989)3287.
\bibitem{s16} G.T.Bodwin,Phys.rev.D31(1985)2616.
\bibitem{s17} S.Gavin and J.Milana,Phys.Rev.lett.,68(1992)1834.
\bibitem{s18} M.B.Johnson et al.,Phys.Rev.,C65,025203(2002).\\
              M.B.Johnson,B.Z.Kopeliovich,I.K.Potashnikova,et.al.Phys.Rev.Lett.,86(2001)4483.
\bibitem{s19} Francois Arleo,Phys.Lett.B532(2002)231.
\bibitem{s20} C.G.Duan L.H.Song,L.J.Huo and G.L.Li,Eur.Phys.J.C29(2003)557.\\
              Duan Chungui,WANG Hong-Min£¬LI Guang-Lie, Chin.Phys.Lett.,
              19(2002)485.
\bibitem{s21} J.Badier.et.al.Nucl.Phys.B484(1997)265;hep-ph/9804212.
\bibitem{s22} G.T.Garvey and J.C.Peng ,Phys.Rev.Lett.,90(2003)092302.
\bibitem{s23} D.Geesaman et. al.,Fermolab Proposal No.E906,1999.
\bibitem{s24}  J.C.Peng et. al.,hep-ph/0007341;M.Asakawa et.al.,KEK Report No 2000-11.
\bibitem{s25}  J.G.Morfin,J.Phys.G29(2003)1935.
               S.A.Kulagin,hep-ph/9812532.
\bibitem{s26}  S.Kumano,hep-ph/0310166.

\end{thebibliography}
\end{document}